\documentclass[twocolumn,showpacs,preprintnumbers,amsmath,amssymb,superscriptaddress,aps]{revtex4}

\usepackage{graphicx}
\usepackage{dcolumn}
\usepackage{bm}
\usepackage[caption=false]{subfig}
\usepackage{epsfig}
\usepackage{graphics}
\usepackage{mathtext}

\begin{document}

\title{Degree Distribution, Rank-size Distribution, and Leadership Persistence in
Mediation-Driven Attachment Networks}

\author{Md. Kamrul Hassan}
\author{Liana Islam}
\affiliation{Department of Physics, University of Dhaka, Dhaka-1000, Bangladesh.}

\author{Syed Arefinul Haque}
\affiliation{Network Science Institute, College of Science, Northeastern University, Boston,
MA-02115, USA.}
\affiliation{Department of Computer Science and Engineering, United International University,
Dhaka-1209, Bangladesh.}

\begin{abstract}%
We investigate the growth of a class of networks in which a new node first picks a mediator
at random and connects with $m$ randomly chosen neighbors of the mediator at each time step.
We show that degree distribution in such a mediation-driven attachment (MDA) network exhibits power-law  
$P(k)\sim k^{-\gamma(m)}$ with a spectrum of exponents depending on $m$. To
appreciate the contrast between MDA and Barab\'{a}si-Albert (BA) networks, we then discuss
their rank-size distribution. To quantify how long a leader, the node with the maximum degree,
persists in its leadership as the network evolves, we investigate the leadership persistence
probability $F(\tau)$ i.e. the probability that a leader retains its leadership up to time
$\tau$. We find that it exhibits a power-law $F(\tau)\sim \tau^{-\theta(m)}$ with persistence
exponent $\theta(m) \approx 1.51 \ \forall \ m$ in the MDA networks and
$\theta(m) \rightarrow 1.53$ exponentially with $m$ in the BA networks.
\end{abstract}

\pacs{61.43.Hv, 64.60.Ht, 68.03.Fg, 82.70.Dd}

\maketitle

\section{Introduction} 

In the recent past, we have amassed a bewildering amount of information on our universe, and
yet we are far from developing a holistic idea. This is because most of the natural and
man-made real-world systems that we see around us are intricately wired and seemingly complex. Much of these
complex systems can be mapped as an interwoven web of large network if the constituents are regarded as nodes or 
vertices and the interactions between constituents as links or edges. 
For example, cells of living systems are networks of molecules linked by chemical interaction, brain
is a network of neurons linked by axons,
the Internet is a network of routers and computers linked by cables or wireless connections, the power-grid
 is a network of substations linked by transmission lines, the World Wide Web (WWW) is a network of
HTML document connected URL addresses \cite{ref.internet, ref.www,
ref.brain, ref.protein}.
Equally, there are many variants of social networks where individuals are nodes or vertices linked 
by social intereactions like friendships, professional ties, there are citation network
where articles are nodes linked by the corresponding citation etc. \cite{ref.coauthorship, ref.movieactor}. 
The first theoretical attempt to guide our understanding about complex network topology began with the seminal 
work of two Hungerians, Paul Erd\"{o}s and Alfr\'{e}d R\'{e}nyi, 
in 1959. The main result of the  Erd\"{o}s-R\'{e}nyi (ER) model 
is that the degree distribution $P(k)$, the probability that a randomly chosen node is connected to $k$ other nodes by one edge,  
is Poissonian revealing that it is almost impossible to find nodes that have
significantly higher or fewer links than the average degree. However, real networks are neither 
completely regular where all the nodes have the same degree $k$ nor completely random where the degree distribution is
Poissonian.

While studying real-life data of some of these systems, 
Barab\'{a}si and Albert found
that the tail of the degree distribution $P(k)$, the probability that a randomly chosen
node is connected to $k$ other nodes, always follows a power-law. In an attempt to explain 
this they realized that real networks are not static, rather they grow with time by continuous addition of new nodes.
They further argued that the new nodes 
establish links to the well connected existing ones {\it preferentially} rather than {\it randomly} - known as the 
preferential attachment (PA) rule. It essentially embodies the intuitive idea of the {\it rich get richer} 
principle of the Matthew effect in sociology \cite{ref.barabasi}. 
Incorporating both concepts, they proposed a model and showed that networks thus grown exhibit power-law
degree distribution $P(k)\sim k^{-\gamma}$ with $\gamma=3$ \cite{ref.barabasi_1, ref.review_1}. Their
findings resulted in a paradigm shift, yet we are compelled to note a couple of drawbacks.
Firstly, the PA rule of the Barab\'{a}si-Albert (BA) model is too
direct in the sense that it requires each new node to know the degree of every node in the
entire network. Networks of scientific interest are often large, hence it is unreasonable to
expect that new nodes join the network with such global knowledge. Secondly, the exponent of
the degree distribution assumes a constant value $\gamma=3$ independent of $m$ 
while most natural and man-made
networks have exponents $2\leq \gamma\leq 3$. In order to avoid this drawback and to provide
models that describe complex systems in more detail, a few variants of the BA model and other
models were proposed. Examples feature mechanisms like rewiring, aging, ranking, redirection,
vertex copying, duplication etc. \cite{ref.rewiring_1, ref.rewiring_2, ref.aging,
ref.ranking, ref.redirection, ref.copying, ref.duplication}. Recently, we have shown that random sequential partition of a square into
contiguous and non-overlapping blocks can be described as a network with power-law degree distribution 
if blocks are regarded as nodes and common border between blocks as links \cite{ref.hassan_njp, ref.hassan_conf}.

\begin{figure}
\centering
\subfloat[]
{
\includegraphics[height=1.4in, width=1.4in, trim = 42mm 42mm 32mm 32mm, clip=true]
{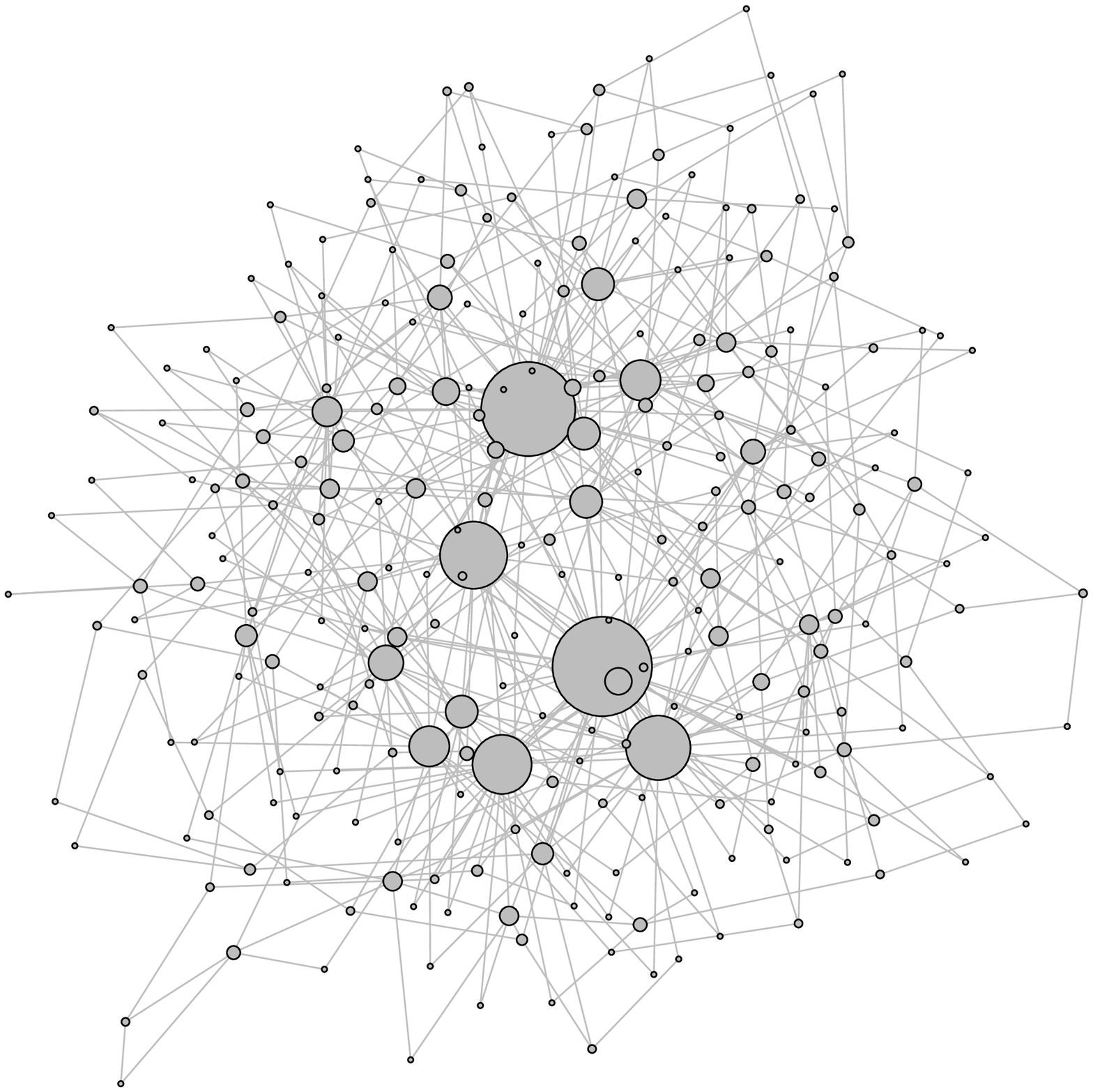}
}
\subfloat[]
{
\includegraphics[height=1.4in, width=1.4in, trim = 42mm 42mm 32mm 32mm, clip=true]
{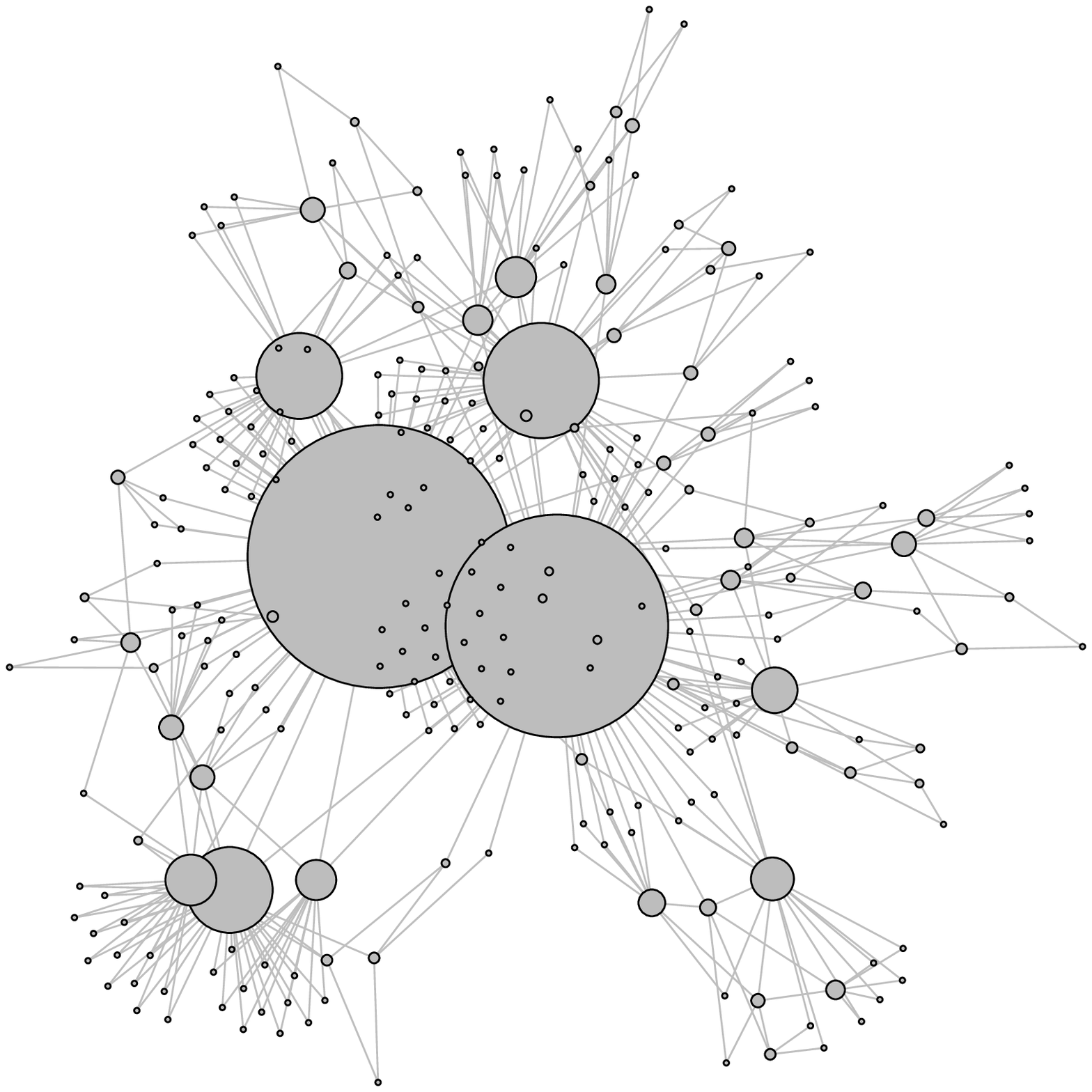}
}
\caption{Visual difference between networks grown by (a) the PA rule of the BA model and (b)
the MDA rule ($m=2$ in both cases). The size of each node is depicted as being proportional
to its degree.} \label{fig:ab}
\end{figure}

In this article, we present a model in which an incoming node randomly chooses an already
connected node, and then connects itself not with that one but with $m$ neighbors of the
chosen node at random. This idea is reminiscent of the growth of the weighted planar
stochastic lattice (WPSL) in whose dual, existing nodes gain links only if one of its
neighbor is picked. Seeing that the dual
of WPSL emerges as a network with power-law degree distribution, we became curious as to what
happens if a graph is grown following a similar rule \cite{ref.hassan_njp, ref.hassan_conf}.
We call it the mediation-driven attachment (MDA) rule since the node that has been picked at
random from all the already connected nodes acts as a mediator for connection between its
neighbor and the new node. Such a rule can embody the preferential attachment process since
an already well-connected node has more mediators and through the mediated attachment
process, it can gain even more neighbors. Finding out the extent of preference in comparison
to the PA rule of the BA model forms an important proposition of this work. There exists a
host of networks where the presence of the MDA rule is too obvious. For instance, while
uploading a document to a website or writing a paper, we usually find documents to link with
or papers to cite through mediators. In social networks such as friendships, co-authorship,
Facebook, and movie actor networks, people get to know each other through a mediator or
through a common neighbor. Thus the MDA model can be a good candidate for describing social
networks.

For small values of $m$, the visual contrast (see FIG. \ref{fig:ab}) of networks grown by the
MDA rule with those grown by the PA rule of the BA model is awe-inspiring. Here we report
that the MDA rule for small $m$ is rather super-preferential as it gives rise to {\it winners
take all} (WTA) effect, where the hubs are regarded as the winners. The scenario is
reminiscent of networks grown by the enhanced redirection model presented by Gabel {\it et
al.} where multiple macrohubs have been observed \cite{ref.gabel_1, ref.gabel_2}. However,
for large $m$, the MDA growth rule increasingly acts more like the simple PA rule as the WTA
effect is replaced by {\it winners take some} (WTS) effect. We solve the model analytically
using mean-field approximation (MFA) and show that the degree distribution exhibits power-law
$P(k)\sim k^{-\gamma}$ with a spectrum of exponents $2\leq \gamma\leq 3$ depending on $m$ such that
$\gamma \rightarrow 3$, when $m$ is large. We perform extensive Monte Carlo simulation to
verify the results. One of the characteristic features of scale-free networks is the
existence of hubs which are linked to an exceptionally large number of nodes. The nodes in
the network can be ranked according to the size of their degree. We show that the rank-size
distribution can provide insights into the nature of the network. Besides, we regard the
richest of all the hubs, the one with the maximum degree, as the leader of the network. In
the context of network theory, we, for the first time, investigate the leadership
persistence probability $F(\tau)$, that a leader retains its leadership up to the time
$\tau$, and find it to decay following power-law with a non-trivial persistence exponent
$\theta(m)$. In particular, $\theta(m)\approx 1.51$ independently of $m$ for MDA networks and for BA
networks it approaches a constant value of $1.53$ exponentially with $m$.

\section{The model}

\begin{figure}
\includegraphics[width=3.6cm,height=3.6cm, trim = 18mm 20mm 20mm 19mm, clip=true]{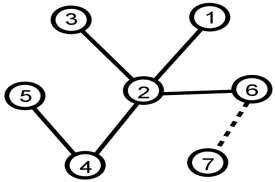}
\caption{A schematic depiction of the MDA rule.} \label{fig2}
\end{figure}

The growth of a network starts from a seed which is defined as a minature network consisting
of $m_0$ nodes already connected in an arbitrary fashion. Below we give an exact algorithm of the
model because we think an algorithm can provide a
better description of the model than the mere definition. It goes as follows:
\begin{enumerate}
 \item[i] Choose an already connected node at random with uniform probability and regard it as
 the mediator.
 
 \item[ii] Pick $m$ of its neighbors also at random with uniform probability.
 
 \item[iii] Connect the $m$ edges of the new node with the $m$ neighbors of the mediator.
 
 \item[iv] Increase time by one unit.
 
 \item[v] Repeat steps (i)-(iv) till the desired network size is achieved.
\end{enumerate}
To illustrate the MDA rule, we consider a seed of $m_0=6$ nodes labeled $i=1,2,...,6$ (FIG.
\ref{fig2}). Now the question is, what is the probability $\Pi(i)$ that an already connected
node $i$ is finally picked and the new node connects with it? Say, the node $i$ has degree
$k_i$ and its $k_i$ neighbors, labeled $1,2,\hdots ,k_i$, have degrees $k_1,k_2,...,k_{k_i}$
respectively. We can reach the node $i$ from each of these $k_i$ nodes with probabilities
inverse of their respective degrees, and each of the $k_i$ nodes can be picked at random with
probability $1/N$. We can therefore write
\begin{equation} \label{eq:1}
\Pi(i)= \frac{1}{N} \Big [ \frac{1}{k_1}+ \frac{1}{k_2} + \hdots + \frac{1}{k_{k_i}} \Big ] =
\frac{\sum_{j=1}^{k_i}{\frac{1}{k_j}}}{N}.
\end{equation}
Numerical simulation suggests that the probabilities $\Pi(i)$ are always normalized, i.e.
$\sum_{i=1}^N\sum_{j=1}^{k_i}{\frac{1}{k_j}}=N$.

The basic idea of the MDA rule is not completely new as either this or models similar to this
can be found in a few earlier works, albeit their approach, ensuing analysis and their
results are different from ours. For instance, Saram\"{a}ki and Kaski presented a random-walk
based model where a walker first lands on an already connected node at random and then takes
a random walk. The new node then connects itself to the node where the walker finally reaches
after $l$ steps \cite{ref.mda_2}. Our model will look exactly the same as the
Saram\"{a}ki-Kaski (SK) model only if the new node arrives with $m=1$ edge and the walker
takes $l=1$ random step. However, they proved that the expression for the attachment
probability of their  model is exactly the same as that of the BA model independently of the
value of $m$ and $l$. Our exact expression for the attachment probability given by Eq.
(\ref{eq:1}) suggests that it does not coincide with the BA model at all, especially for
$m=1$ case. On the other hand, the model proposed by Boccaletti {\it et al.} may appear similar to
ours, but it markedly differs on closer look. The incoming nodes in their model has the
option of connecting to the mediator along with its neighbors
\cite{ref.boccaletti}. Nevertheless, in has been shown in both \cite{ref.mda_2} and
\cite{ref.boccaletti} that the exponent $\gamma=3$ independent of the value of $m$, which is
again far from what our model entails.

As far as the definition of the MDA model is concerned, it is exactly the same as the
one recently studied by Yang {\it et al.}. They too gave a form for $\Pi(i)$ and resorted to
mean-field approximation \cite{ref.mda_1}. However, the nature of their expressions are
significantly different from ours, and we have justified our version of the mean-field
approximation on a deeper level by drawing conclusions from our study of the inverse of the
harmonic mean of degrees of the neighbors of each and every node in our simulated networks.
We shall see below that their results, both by mean-field approximation and numerical
simulation, do not agree with our findings. Yet another closely related model is the Growing
Network with Redirection (GNR) model presented by Gabel, Krapivsky and Redner where at each time
step a new node either attaches to a randomly chosen target node with probability $1-r$, or
to the parent of the target with probability $r$ \cite{ref.redirection}. The GNR model with
$r=1$ may appear similar to our model. However,  it should be noted that unlike the GNR
model, the MDA model is for undirected networks, and that the new link can connect with any
neighbor of the mediator- parent or not. One more difference is that, in our model new node may
join the existing network with $m$ edges and they considered $m=1$ case only. We shall show that 
the role of $m$ in this model is crucial. 

\section{Mean-field approximation}

\begin{figure}[htb]
\begin{center}
\includegraphics[width=5.0cm,height=8.5cm,clip=true,angle=-90]{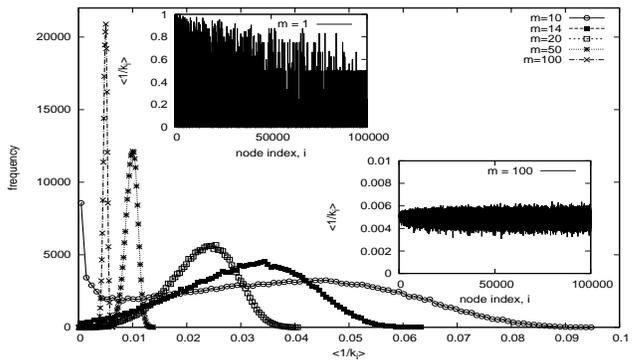}
\caption{The distribution of IHM for different $m$ values. The fluctuations of IHM of individual
nodes is shown in the inset (top left for $m=1$ and bottom right for $m=100$).}
\label{fig:histogram_inset_fluctuations}
\end{center} \label{fig3}
\end{figure}

\begin{figure}[htb]
\begin{center}
\includegraphics[width=5.5cm,height=8.0cm,clip=true,angle=-90]{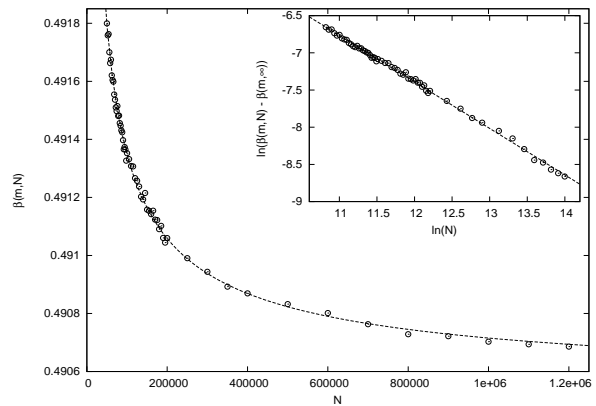}
\caption{Plot of $m\overline{\Big <{\frac{1}{k_i}}\Big >}\equiv \beta(m,N)$ vs. $N$ for $m=70$,
and inset showing that $\beta(m,N)$ saturates to $\beta(m,\infty) = 0.490513$ algebraically as
$N^{-0.62516}$.} \label{fig:m70_beta_approximate_vs_N_verification_of_law}
\end{center} \label{fig4}
\end{figure}

The rate at which an arbitrary node $i$ gains links is given by
\begin{equation} \label{eq:2}
{\frac{\partial k_i}{\partial t}} = m \ \Pi(i)
\end{equation}
The factor $m$ takes care of the fact that any of the $m$ links of the newcomer may connect
with the node $i$. Solving Eq. \eqref{eq:2} when $\Pi(i)$ is given by Eq. \eqref{eq:1} seems
quite a formidable task unless we can simplify it. To this end, we find it convenient to
re-write Eq. \eqref{eq:1} as
\begin{equation} \label{eq:npi}
\ \Pi(i) = \frac{k_i}{N} \ \frac{\sum_{j=1}^{k_i}{\frac{1}{k_j}}}{k_i}.
\end{equation}
The factor $\frac{\sum_{j=1}^{k_i}{\frac{1}{k_j}}}{k_i}$ is the inverse of the harmonic mean
(IHM) of degrees of the $k_i$ neighbors of a node $i$. We performed extensive numerical
simulation to find out the nature of the IHM value of each node. We
find that for small $m$ (roughly $m<14$), their values fluctuate so wildly that the
mean of the IHM values over the entire network bears no meaning. It can be seen from
the frequency distributions of the IHM values shown in FIG. \ref{fig:histogram_inset_fluctuations}.
However, as $m$ increases beyond $m=14$ the frequency distributions
gradually become more symmetric and the fluctuations occur at increasingly lesser extents,
and hence the meaning of the mean IHM seems to be more meaningful. Also interesting is the fact
that for a given $m$ we find that the mean IHM value in the large $N$ limit becomes constant
which is demonstrated in FIG. \ref{fig:m70_beta_approximate_vs_N_verification_of_law}. 

The two factors that the mean of the IHM is meaningful and it is independent of $N$ 
implies that we can apply the mean-field approximation (MFA). That is, within this approximation
 we can replace the true
IHM value of each node by their mean.  In this way, all the information on
correlations in the fluctuations is lost. The good thing is that it makes the problem analytically tractable.
One immediate consequence of it is that the MDA rule with $m>14$, like the BA model, is
preferential in character since we get $\Pi(i)\propto k_i$. It implies that the higher the
links (degree) a node has, the higher its chance of gaining more links since they can be
reached in a larger number of ways through mediators which essentially embodies the intuitive
idea of rich get richer mechanism. Therefore, the MDA network can be seen to follow
the PA rule but in disguise. Moreover, for small $m$ the MFA is no longer valid. We shall see
below that for small $m$ the attachment
probability is in fact superpreferential in character.

\section{Degree and Rank-Size Distribution}

\begin{figure}[htb]
\begin{center}
\includegraphics[width=4.8cm,height=8.0cm,clip=true,angle=-90]{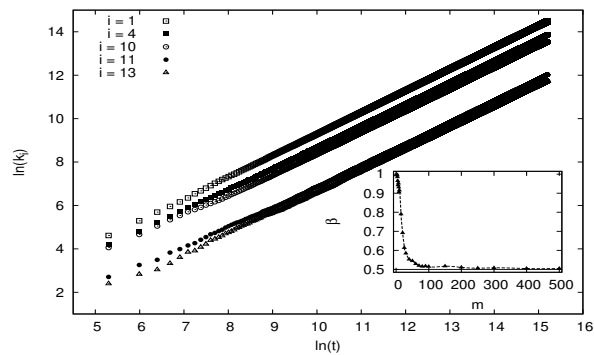}
\caption{Plots of $\ln(k_i)$ versus $\ln(t)$ for $5$ nodes added at five different times for
$m = 2$. In the inset, we show variation of the slope $\beta$ with $m$.} \label{fig:kth2.eps}
\end{center} \label{fig5}
\end{figure}

It is noteworthy that the size $N$ of the network is an indicative of time $t$ since we assume
that only one node joins the network at each time step. Thus, for $N>>m_0$ we can write
$N \sim t$. Using this, Eq. \eqref{eq:npi}, and MFA in Eq. \eqref{eq:2}, we find the rate
equation in a form that is easy to solve, which is
\begin{equation} \label{eq:3}
{\frac{\partial k_i}{\partial t}} = k_i \ \frac{\beta (m)}{t}.
\end{equation}
Here we assumed that the mean IHM is equal to $\beta(m)/m$ for large $m$ where the factor $m$ in the
denominator is introduced for future convenience. Solving the above rate equation subject to the initial condition that the $i$th node is born at time $t=t_i$ with
$k_i(t_i)=m$ gives,
\begin{equation} \label{eq:4}
 k_i(t) = m \Big ({\frac{t}{t_i}}\Big )^{\beta(m)}.
\end{equation}
As the FIG. (\ref{fig:kth2.eps}) shows, the numerical results are in good agreement with this
growth law for $k_i(t)$. Note that the solution for $k_i(t)$ is exactly the same as that of the BA model except the fact that the numerical value of
the exponent $\beta(m)$ is different and depends on $m$. We, therefore, can immediately write
the solution for the degree distribution
\begin{equation} \label{eq:20}
 P(k) \sim k^{-\gamma(m)}, \: \text{where} \: \: \gamma(m) = \frac{1}{\beta(m)}+1.
\end{equation}
The most immediate difference of this result from that of the BA model is that the exponent
$\gamma$ depends on $m$. 

\begin{figure}[htb]
\begin{center}
\includegraphics[width=5.0cm,height=8.5cm,clip=true,angle=-90]{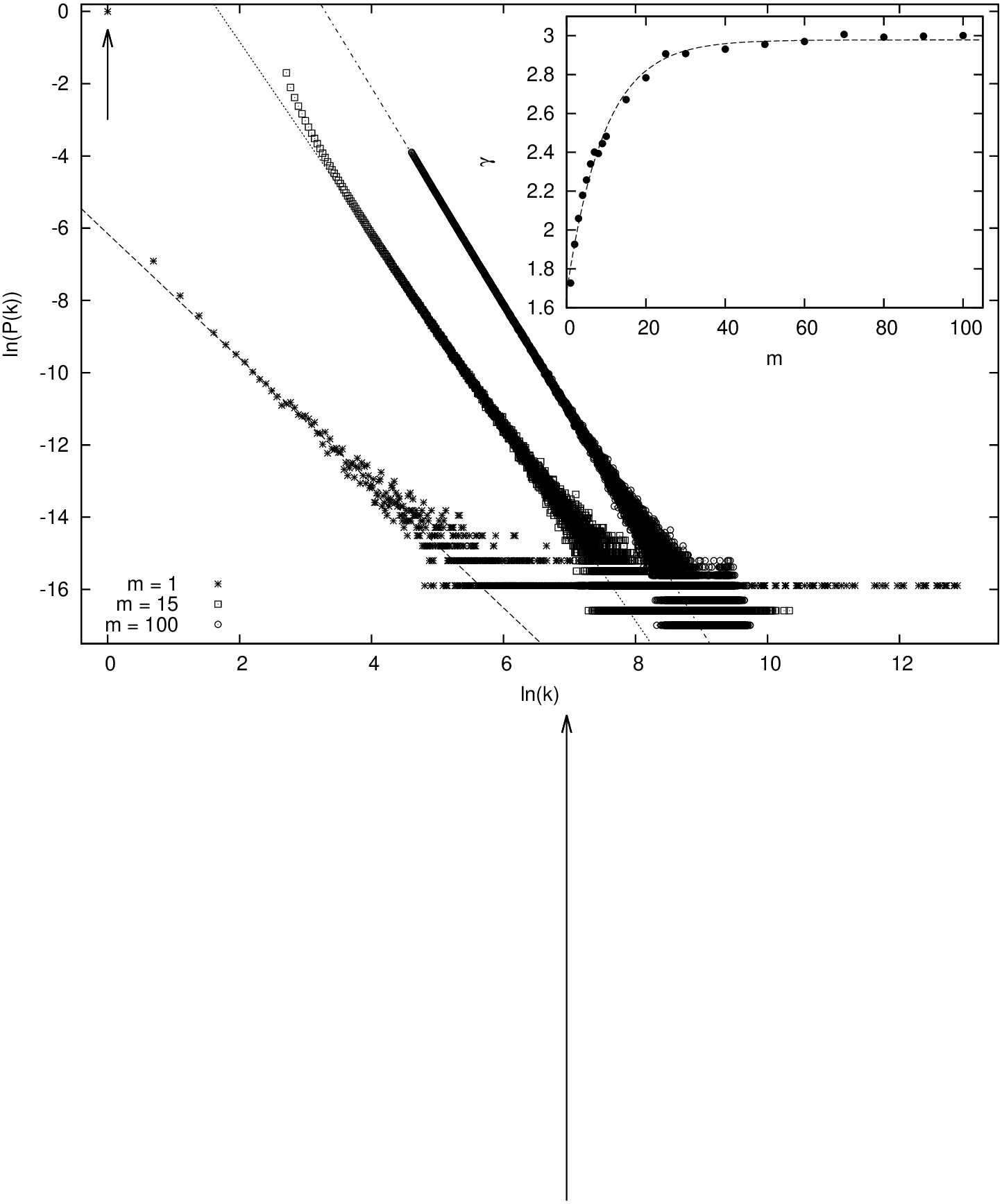}
\caption{Plots of $\ln (P(k))$ vs $\ln(k)$ for $m = 1$, $m=15$ and $m=100$ revealing that the
MDA rule gives rise to power-law degree distribution. In the inset we show the variation in the
exponent $\gamma$ as a function of $m$.}\label{fig:deg_dist_gm}
\end{center} \label{fig6}
\end{figure}

To verify Eq. (\ref{eq:20}) we plot $\ln(P(k))$ vs. $\ln(k)$ in FIG.
(\ref{fig:deg_dist_gm}) using data extracted from numerical simulation. In all cases, we
find straight lines with characteristic fat-tail which confirms that our analytical solution
is in agreement with the numerical solution. It is important to note that, for small $m$,
especially for $m=1$ and $2$, there is one point in each of the $\ln(P(k))$ vs. $\ln(k)$
plots that stands out alone. These special points correspond to $P(k=1)\approx 99.5\%$ for
$m=1$, and $P(k=2)\approx 95.4\%$ for $m=2$. This implies that almost $99.5\%$ and $95.4\%$
of all already connected nodes are held together by a few hubs and super-hubs for $m=1$ and
$2$ respectively. This huge percentage of nodes are minimally connected and this embodies the
intuitive idea of WTA phenomenon. We find that as $m$ increases, this percentage decreases
and roughly at about $m=14$, all the data points follow the same smooth trend revealing a
transition from the WTA effect to the WTS effect. This happens when the IHM has less noise
and the mean of IHM becomes increasingly more meaningful. The IHM value can thus be regarded
as a measure of the extent up to which the degree distribution follows power-law. In their
GNR model with enhanced redirection mechanisms, Gabel, Krapivsky and Redner also noticed such
highly dispersed networks which tend to be dominated by one or a few high-degree nodes
\cite{ref.gabel_1, ref.gabel_2}.

\begin{figure}[htb]
\begin{center}
\includegraphics[width=5.0cm,height=8.5cm,clip=true,angle=-90]{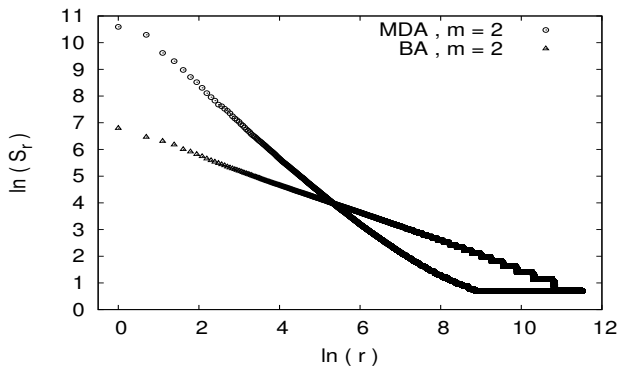}
\caption{To be able to appreciate the contrast between the MDA and BA network we plot the
rank-size distribution $\ln(S_r)$ vs $\ln(r)$ for $m=2$.
} 
\label{fig:rank_size}
\end{center}
\end{figure}

It is interesting to note that the exponent $\gamma$ of the degree distribution increases
with increasing $m$. We used gnuplot to fit the function $g(x) = P(1-Q e^{-R x})$ to the 
$\gamma$ versus $m$ data and find that the data satisfies the relation
$\gamma(m) \approx 2.97457(1 - 0.449478 e^{-0.110688m})$ quite well. The asymptotic standard errors in
the values of $P$, $Q$, and $R$ are $\pm 0.01757$, $\pm 0.01407$, and $\pm 0.007378$
respectively. This result stands in sharp contrast with that of Yang {\it et al.} who found
the exponent $\gamma$ to vary in the range from $1.90$ to $2.61$ \cite{ref.mda_1}. Also, for
$m=1$, which is the case closest to $m=1$ and $l=1$ of the SK model, we found that
$\gamma=1.65480$ and not $3$. On the other hand, in the large $m$ limit we find $\gamma \rightarrow 3$. It
hints at a possible similarity with the BA model where $\gamma =3$. To find the extent
of similarity or dissimilarity of our model with the BA model, we tuned the values of $m$ and
looked at the rank-size distribution. The rank-size distribution can be found to describe
remarkable regularity in many phenomena including the distribution of city sizes, the sizes
of businesses, the frequencies of word usage, and wealth among individuals
\cite{ref.power_law_newman}. These are all real-world observations where the rank-size
distributions follow power-laws. In our case, we measured the size of each node by its
degree, and the nodes which have the largest degree are given rank $1$, the nodes which have
the second-largest degree rank $2$, and so on. Assuming $S_r$ to be the size of the nodes
having rank $r$ we plotted $\log (S_r)$ vs $\log (r)$ in FIG. \ref{fig:rank_size}. It clearly
reveals that the size distribution of nodes decays with rank following the Pareto law. Note
that, the power-law distribution pattern for the rank-size is often found to be true only
when very small sizes are excluded from the sample. However, our main focus is on the
difference between the BA and the MDA model. Clearly, for small $m$, the size of the hubs of
an MDA ($S_r$ with small $r$) network is far more rich than that of a BA network. However,
for large $m$, the rank distributions of the two networks become almost identical and this is
consistent with the fact that the exponent $\gamma(m)$ of the MDA network in the large $m$
limit coincides with that of the BA network.

\section{Leadership Persistence}

In the growing network not all nodes are equally important. The extent of their importance is
measured by the value of their degree $k$. Nodes which are linked to an unusually large
number of other nodes, i.e. nodes with exceptionally high $k$ value, are known as hubs. They are
special because their existence make the mean distance, measured in units of the number of
links, between nodes incredibly small thereby playing the key role in spreading rumors,
opinions, diseases, computer viruses etc. \cite{ref.epidemic}. It is, therefore, important to
know the properties of the largest hub, which we regard as the leader. Like in society, the
leadership in a growing network is not permanent. That is, once a node becomes the leader, it
does not mean that it remains the leader {\it ad infinitum}. An interesting question is: how
long does the leader retain this leadership property as the network evolves? To find an
answer to this question, we define the leadership persistence probability $F(\tau)$ that a
leader retains its leadership for at least up to time $\tau$. Persistence probability has
been of interest in many different systems ranging from coarsening dynamics to fluctuating
interfaces or polymer chains \cite{ref.persistence_bray_majumdar}.

\begin{figure}[htb]
\begin{center}
\includegraphics[width=7.0cm,height=8.5cm,clip=true,angle=-90]{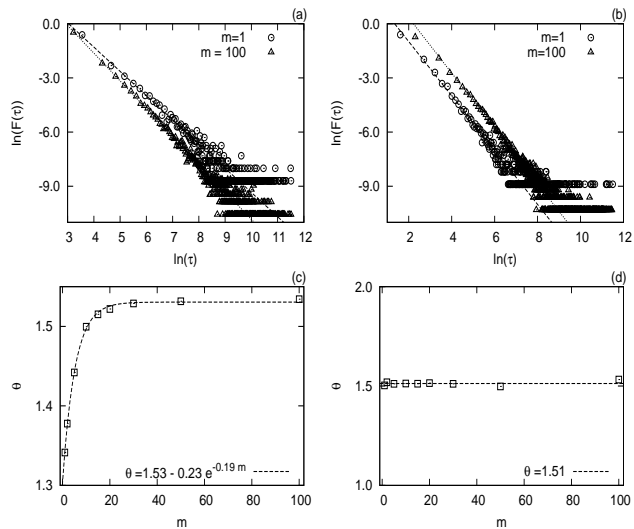}
\caption{The two plots at the top reveal the power-law behaviour of the leadership
persistence probability. To appreciate the role of $m$ we give leadership persistence
probability for two values ($m=1$ and $m=100$) in the same plot: (a) BA networks and (b) MDA
networks. The two plots at the bottom are for the persistence exponent $\theta$ as a function
of $m$ in (c) BA networks and (d) MDA networks.} \label{fig:persistenc}
\end{center}
\end{figure}

We find it worthwhile to look into the leadership persistence probability $F(\tau)$ first in
MDA networks and then in BA networks to see their difference. We perform $M=500$ independent
realizations under the same condition and take records of the duration time $\tau$ in which
leadership has not changed. The leadership persistence probability $F(\tau)$ is then obtained
by finding the relative frequency of leadership change out of $M$ independent realizations
within time $\tau$. The plots in (a) and (b) of FIG. (\ref{fig:persistenc}), show
$\log(F(\tau))$ as a function of $\log(\tau)$ for BA and MDA networks respectively. In each
case, we have shown plots for $m=1$ and $m=100$ to be able to appreciate the role of $m$ in
determining the persistence exponent $\theta(m)$. These plots for different $m$ result in
straight lines revealing that persistence probability decays as
\begin{equation}
F(\tau)\sim \tau^{-\theta(m)}.
\end{equation}
One of the characteristic features of $F(\tau)$ is that like $P(k)$, it too has a long tail
with scarce data points that results in a "fat" or "messy" tail when we plot it in the
log-log scale. The persistence exponent $\theta$ carries interesting and useful information
about the full history of the dynamics of the system. Therefore, its prediction is of
particular importance. We find that the exponent $\theta$ for MDA networks is almost equal to
$1.51$ independently of $m$. On the other hand, in the case of BA networks, it rises to
$1.53$ exponentially with $m$ i.e., it depends on $m$. This is just the opposite of what
happens to the exponent $\gamma$ of the degree distribution. In many natural and man-made
networks, the leadership is a stochastic variable as it varies with time. For instance, the
leader in the World Wide Web is not static, and hence it would be interesting to see the
leadership persistence properties there. One could also look at such persistence properties
in other growing networks.

\section{Conclusion}

In this paper, we have elaborated on a mediation driven attachment rule for growing networks
that exhibit power-law degree distribution. At a glance, it may seem that it defies the PA
rule that Barab\'{a}si and Albert argued to be an essential ingredient for the emergence of
power-law degree distribution. However, we have shown explicitly that the MDA rule is in fact
not only preferential but also super-preferential in some cases. In most cases, it embodies
at least the intuitive idea of the PA rule, albeit in disguise. We obtained an exact
expression for the mediation-driven attachment probability $\Pi(i)$ with which a new node
picks an already connected node $i$. Solving the model analytically for the exact expression
of $\Pi(i)$ appeared to be a formidable task. However, the good news is that we could still
find a way to solve it using mean-field approximation. Later, it turned out that not being
able to solve analytically for the exact expression for $\Pi(i)$ was highly rewarding as it
helped us gain deeper insight into the problem. While working with the expression for
$\Pi(i)$, we realized that the IHM value plays a crucial role in the model. 

We have found that the fluctuations in the IHM values of the existing nodes are so wild for
small $m$, that the mean bears no meaning and hence mean-field approximation is not valid.
However, for large $m$, the fluctuations get weaker and their distribution starts to peak
around the mean, revealing that the mean has a meaning. This is the regime where MFA works
well and we verified it numerically. Here we found that the WTA phenomenon is replaced by the
WTS phenomenon. Besides, we all know that the leader of the hubs is the most important of all
nodes in the sense that it is the most connected. We investigated the leadership persistence
probability in both BA and MDA networks and found that it decays following a power-law. We
found that the persistence exponent $\theta\approx 1.51$ is independent of $m$ for MDA
networks and grows to a constant $\theta\approx 1.53$ exponentially with $m$ for BA networks.
We hope to extend our work to study dynamic scaling and universality classes in networks
grown using the MDA model for different $m$, and check how they differ from similar aspects
in the BA model \cite{ref.mda_data_collapse}.

\end{document}